\documentclass[pra,twocolumn,graphicx]{revtex4}
\usepackage{epsfig,amsmath}

\begin{document}

\title{Nonsequential double ionization with few-cycle laser pulses}
\author{X. Liu}
\affiliation{Max-Born-Institut f\"{u}r nichtlineare Optik und Kurzzeitspektroskopie,\\
Max-Born-Str. 2A, D-12489 Berlin, Germany}
\author{C. Figueira de Morisson Faria}
\affiliation{Institut f\"{u}r theoretische Physik, Universit\"{a}t Hannover, Appelstr. 2,
30167 Hannover}
\date{\today}

\begin{abstract}
We investigate differential electron momentum distributions in
non-sequential double ionization (NSDI) with linearly polarized, few-cycle
pulses, using a classical model based on a laser-assisted inelastic
$(e^-,2e^-)$ rescattering 
mechanism. We show that these yields, as
functions of the momentum components
parallel to the laser polarization, are highly asymmetric and strongly
influenced by the absolute phase, i.e., the
phase difference between the carrier oscillation of a few-cycle pulse and
its envelope. Indeed, around a critical phase, such distributions 
 change their sign in a
radical fashion. 
This phase dependence provides a possibility for absolute-phase measurements which is, in principle, superior to the schemes involving 
high-order harmonic generation  or
above-threshold ionization. 
\end{abstract}

\maketitle

 Recently, few-cycle laser pulses of intensities around or higher than $10^{14}\mathrm{%
W/cm}^{2}$ have proven to be extremely important. A
particular characteristic of such pulses is that they may have very high
intensities and, still, carry much less energy than their longer
counterparts, such that, effectively, ionization is reduced. This has
extended the damage threshold of solid-state materials up to the
intensities in question \cite{damage}, and has 
made the generation of high-order harmonic
radiation up to astonishingly high frequencies possible \cite{fewcreview}. Furthermore,
their length, of the order of a few $fs$, permits controling processes such as molecular motion or chemical reactions
 \cite{attocontrol}, as well as the production of isolated, X- ray attosecond
pulses \cite{atto1}.

In this pulse-length regime, the so-called \textquotedblleft absolute
phase\textquotedblright , i.e., the phase difference between the envelope of
the pulse and its carrier frequency, has considerable influence on several
strong-field phenomena, as for instance high-order harmonic
generation (HHG) \cite{fewchhg} and above-threshold ionization \cite{footnati} (ATI) \cite%
{fewcati1,fewcati2}. Indeed, this phase determines features such as the
maximal harmonic or photoelectron energy, the time profile of both
phenomena, and the intensity of the ATI or HHG yields.

This is not surprising, since the physics of HHG and ATI
is directly related to the {\it instantaneous}, time dependent field. HHG,
 for instance, is described by a three-step mechanism in
which an electron leaves an atom at an instant $t_{0}$ through tunneling
ionization, propagates in the continuum under the influence of the external
laser field and, at a later time $t_{1},$ recombines with a bound state of
the parent ion, generating harmonics \cite{corkum}. Slightly different
processes, either in which elastic rescattering with the parent ion is taken
as the third step, or in which the electron reaches the detector without
recolliding explain the high-order and low-order ATI peaks, respectively 
\cite{aticlass}.

In order to interprete the experimental data obtained in such cases, the
precise knowledge of the absolute phase $\phi $ is required. This poses a
serious practical problem, since this phase is difficult to stabilize, to
control or to measure \cite{phasecontrol}. For this reason, schemes for
measuring $\phi $ have been suggested and realized, as,
for instance,
 using the asymmetry in ATI photoelectron counts \cite{fewcati1}.

In this Letter, we propose laser-assisted nonsequential double ionization
(NSDI) as a tool for absolute phase diagnosis. \ This phenomenon is being
the subject of very active discussion, which was triggered by differential
measurements of electron momentum distributions performed with the COLTRIMS
(Cold Target Recoil Ion Momentum Spectroscopy) technique, for linearly
polarized fields of intensities of the order of $I\sim 10^{14}-10^{15}%
\mathrm{W/cm}^{2}$ incident in rare-gas samples \cite{expe1}. Such
measurements revealed very peculiar features, namely two symmetric peaks at $%
p_{1\parallel }=p_{2\parallel }=\pm 2\sqrt{U_{p}},$ in the $\left(
p_{1\parallel },p_{2\parallel }\right) $ plane, where $p_{j\parallel
}(j=1,2) $ and $U_{p}$ denote the momentum components parallel to the laser
field polarization and the ponderomotive energy \cite{footnpond}, respectively.

These features are explained by a physical mechanism very similar to those
in HHG and high-order ATI.
 The main difference lies on the rescattering process at $t_{1},$ which is now
inelastic: the first electron gives part of its kinetic energy upon return
to the second electron, so that it can overcome the ionization potential of
the singly ionized atom and reach the continuum \cite{corkum}.

This laser-assisted rescattering process has been considered by several
groups, using either classical \cite{classical}, semiclassical \cite%
{doublegoresl,nsdiuni,preprint0,preprint} or quantum-mechanical \cite%
{richard,tdse,repulsion} approaches, different types of electron-electron
interaction \cite{doublegoresl,preprint0,preprint}, and neglecting or
including electron-electron repulsion in the final states \cite%
{repulsion,preprint0,preprint}. So far, since the pulses involved were
relatively long, they have been mostly approximated by monochromatic fields.

Particularly what classical models concern, it is astonishing how well they
reproduce the main features either observed experimentally, or obtained by
means of more refined, quantum mechanical methods. Indeed, recently, we have computed NSDI yields considering rescattering in its simplest form, i.e., electron-impact ionization, both 
classically and within a quantum-mechanical S-Matrix framework, with practically identical 
results \cite{preprint0,preprint}.

In this work,  we use a similar
classical model as in \cite{preprint0,preprint}, in which an electron ensemble is
subject to a few-cycle pulse $E(t)=-dA(t)/dt$. The vector potential is given by
\begin{equation}
\mathbf{A}(t)=A_{0}\sin ^{2}(\Omega t/2)\sin (\omega t+\phi )\hat{e}_x,
\end{equation}%
with frequency $\omega $, amplitude $A_{0}$, absolute phase $\phi $, 
and $\Omega =\omega /n$, where $n$ denotes the number of cycles. The
electrons are ejected in the continuum at a time $t_0$ with vanishing drift
velocities and from the origin of the coordinate system. The start times are
uniformly distributed and the ejection probability per unit time, unless 
stated otherwise, is given by the quasi-static 
\cite{quasist} tunneling rate%
\begin{equation}
R(t_{0})\sim \frac{1}{|E(t_{0})|}\exp \left[ \frac{-2(2|E_{01}|)^{3/2}}{%
3|E(t_{0})|}\right] ,  \label{tunelrate}
\end{equation}%
where $|E_{01}|$ is the ionization potential of the atom in question. 
Subsequently, these electrons propagate under the 
influence of only the laser field. Finally, 
some of them return to the site of their release and free a
second ensemble of electrons through inelastic collisions at a
later instant $t_{1}$.

The equations of motion of each pair in such electron ensembles, in atomic
units, read
\begin{equation}
\lbrack \mathbf{k}+\mathbf{A}(t_{0})]^{2}=0,  \label{ioniz}
\end{equation}

\begin{equation}
\int_{t_{0}}^{t_{1}}[\mathbf{k}+\mathbf{A}(t)]^{2}dt=0  \label{return}
\end{equation}
and
\begin{equation}
\sum_{j=1}^{2}[\mathbf{p}_{j}+\mathbf{A}(t_{1})]^{2}=[\mathbf{k}+%
\mathbf{A}(t_{1})]^{2}-2|E_{02}|.  \label{rescatt}
\end{equation}
Eq. (\ref{ioniz}) gives the energy conservation at the ionization time.
 Eq. (\ref{return}) imposes restrictions upon
the intermediate electron momentum \ $\mathbf{k}$ such that the electron
returns to its parent ion. Finally, the third expression (Eq. (\ref{rescatt}%
)) yields the energy conservation at the recollision time $\ t_{1}$. Thereby,
the first electron gives part of its kinetic energy $E_{\mathbf{ret}}(t_1)=[%
\mathbf{k}+\mathbf{A}(t_{1})]^{2}/2$ upon return to the second
electron, so that it is able to overcome the ionization potential $|E_{02}|$ of the singly
ionized atom. In terms of the momentum components parallel and perpendicular
to the electric-field polarization, denoted $p_{j\parallel }$ and $\mathbf{p}%
_{j\perp }$, respectively, Eq. (\ref{rescatt}) is written as%
\begin{equation}
\sum_{j=1}^{2}[p_{j\parallel }+A(t_{1})]^{2}=[\mathbf{k}+\mathbf{A}%
(t_{1})]^{2}-2|E_{02}|-\sum_{j=1}^{2}\mathbf{p}_{j\perp }^{2}.
\label{resc2}
\end{equation}%
Eq. (\ref{resc2}) describes a circle in the $(p_{1\parallel },p_{2\parallel
}) $ plane, centered at $A(t_{1})$ and whose radius depends on $E_{\mathbf{%
ret}} $, $|E_{02}|$ and on $\mathbf{p}_{j\perp }^{2}(j=1,2).$ 
The transverse momenta effectively shift the binding energy
which must be overcome such that, depending on this quantity, there are
situations for which the rescattering process is classically
forbidden.

The electron momentum distributions  then read
\begin{equation}
\Gamma \sim \int dt_{0}R(t_{0})\delta \left( E_{\mathrm{ret}%
}(t_{1})-\sum_{j=1}^{2}\frac{(\mathbf{p}_{j}+\mathbf{A}(t_{1}))^{2}}{%
2}-|E_{02}|\right)  ,\label{yield}
\end{equation}%
where the argument of the $\delta$ function gives the energy conservation upon return. The transverse momenta are integrated over. Details about this model
are given in \cite{preprint}.

 These distributions are
displayed in the upper panels of Fig. 1, for neon \cite{footneon}, as density plots in
the $(p_{1\parallel },p_{2\parallel })$ plane. 
Their circular shapes and the maxima along $p_{1\parallel}=p_{2\parallel}$  are features also present for
monochromatic driving fields \cite{footnint}, and mean, physically, that both electrons are leaving the parent ion most probably with equal parallel momenta. However, the fact that the yields are 
concentrated in only one quadrant of the $(p_{1\parallel
},p_{2\parallel })$ plane, makes them strikingly different from the former distributions,  
 which are symmetric in $(p_{1\parallel
},p_{2\parallel })\leftrightarrow $ $(-p_{1\parallel },-p_{2\parallel })$.
Furthermore, for a narrow phase interval around 
a critical phase $\phi_c$ (c.f. panel (c)), the sign of the
momenta $p_{j\parallel }$ change in a rather radical fashion, and the whole
yield is shifted from the first to the third quadrant. For increasing pulse length, these effects get less pronounced and practically disappear (Figs. 1(e)--1(h)). The distributions then become symmetric and phase-independent.

Important questions concern the physical origin of both the asymmetry and
the critical phase: are they caused by the phase space or by the tunneling
rate (\ref{tunelrate})? Depending on the parameters, a 
whole phase-space region may become classically inacessible, such that the
radius of the circle described by Eq. (\ref{resc2}) would collapse and the
corresponding NSDI yields would vanish. The quasi-static tunneling rate, on
the other hand, favors the start times $t_{0}$ for which the instantaneous
field amplitude $|E(t_{0})|$ is large, as compared to those for which $%
|E(t_{0})|$ is small. Thus, the contributions to the yield 
from the former or from the latter case would be enhanced or
suppressed, respectively.

\begin{figure}
\begin{center}
\includegraphics[width=4.8cm,angle=270]{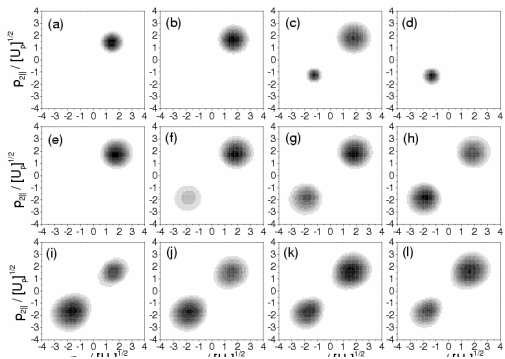}
\end{center}
\caption{Electron momentum distributions along the laser polarization,
for neon ($|E_{01}|=0.79$ a.u. and $|E_{02}|=1.51$ a.u.), subject to pulses of intensity
$I=4.7\times 10^{14}\mathrm{W/cm^2}$ and carrier frequency
$\omega=0.057$ a.u., respectively. In panels (a) to (d) and (i) to (l) we consider a four-cycle pulse ($n=4$), whereas in panels (e) to (h) the 
pulse length is varied.
Panels (a) to (h) and (i) to (l) were computed with the quasi-static and a constant tunneling rate, respectively.
Panels (a) and (i): $\phi=0.1\pi$; panels (b) and (j): $\phi=0.5\pi$; panels
(c) and (k): $\phi=0.8\pi$; panels (d) and (l): $\phi=0.9\pi$. In panels (e), 
(f), and (g) $n=8$, $n=12$ and $n=16$ respectively, and $\phi=0.1\pi$. In panel (h), $n$ is the same as in (g) and $\phi=0.9\pi$.
}
\end{figure}
In order to single out the influence of the phase space, we assume that
the electrons belonging to the first ensemble reach the continuum at a constant rate. Such results
are shown in panels (i)--(l) of Fig. 1, and are radically different from
those obtained with the more realistic, quasi-static tunneling rate. Indeed,
the momentum distributions, though asymmetric, exhibit
peaks in \textit{both} first and third quadrants of the $(p_{1\parallel
},p_{2\parallel })$ plane, vaguely resembling those obtained with
monochromatic driving fields. The asymmetry is expected, since, for such
pulses, the relation $A(t)=-A(t\pm T/2),$ and thus $|\Gamma (t_{1},t_{0},p_{1%
\parallel },p_{2\parallel })|=|\Gamma (t_{1}\pm T/2,t_{0}\pm T/2,-p_{1\parallel
},-p_{2\parallel })|$, where $T=2\pi /\omega ,$ which was true for
monochromatic fields, no longer holds. However, the huge effects observed in
the upper panels
are absent. Physically, this means that
there is a momentum region for which the rescattering process is classically
allowed but for which the probability that the first electron reaches the
continuum is very small. Consequently, even if this region is
large, its contributions to the yield will be negligible.

In Fig. 2 we analyze this effect in detail.  Therein, the electron start times
$t_0$ are plotted, together with the quasi-static rate. We
restrict the parameter range so that the classically allowed region
is most extense \cite{footnpspace}, taking parallel momenta along the diagonal $%
p_{1\parallel }=p_{2\parallel }=p_{\parallel }$ and vanishing transverse
momenta
\cite{nsdiuni}. We consider only \ pairs $%
(t_{1},t_{0})$ of start and return times for the first electron such that
its excursion time $\Delta t=t_{1}-t_{0}$ in the continuum is of the order
of $T/2$ \cite{footnspread}.
\begin{figure}
\begin{center}
\includegraphics[width=4.8cm,angle=270]{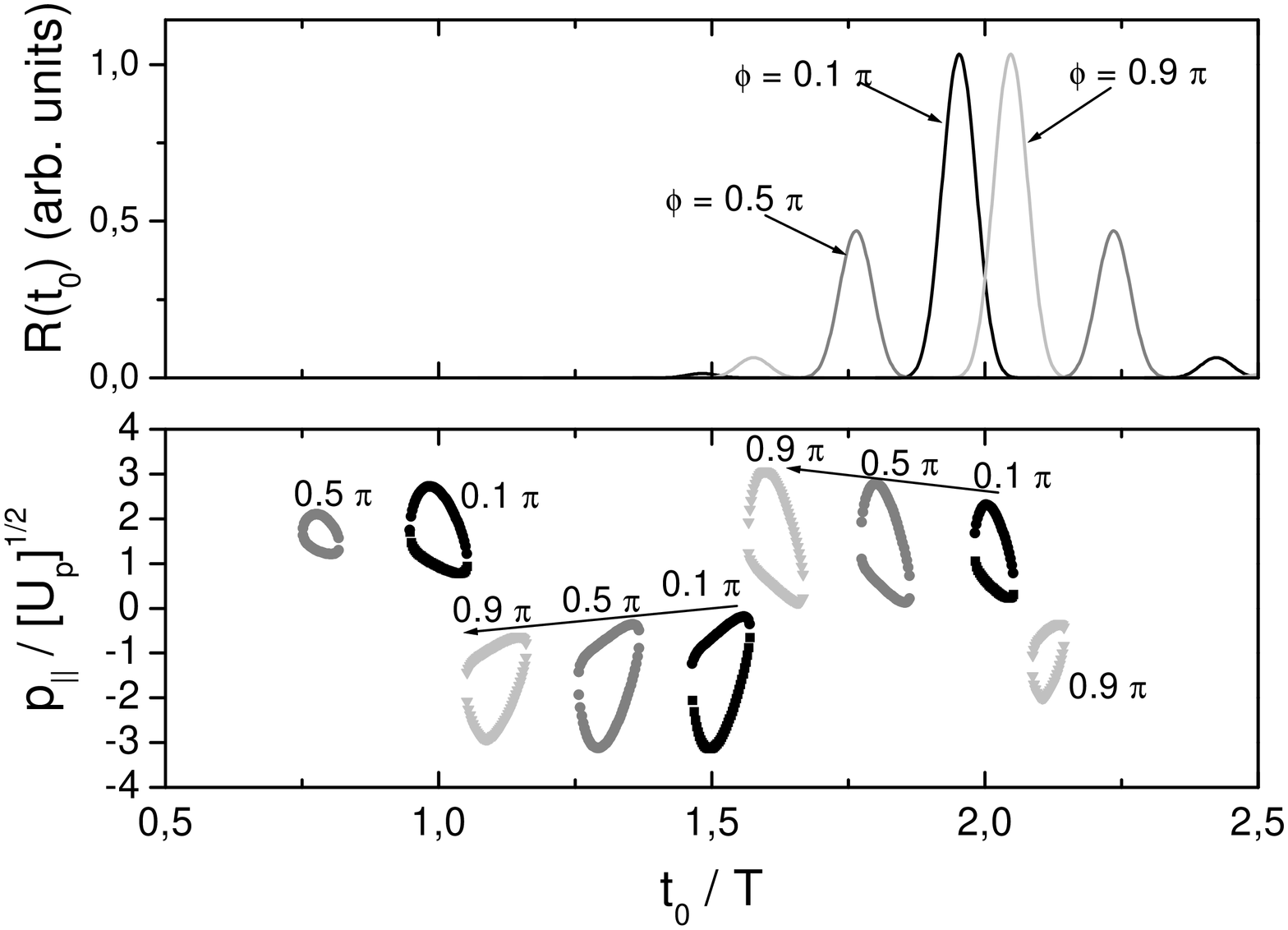}
\end{center}
\caption{ 
Parallel electron momenta $p_{\parallel }$ along  $p_{1\parallel
}=p_{2\parallel }$, for transverse momenta $\mathbf{p}_{1\perp}=\mathbf{p}_{2\perp}=0$, as functions of the start times $t_{0}$ of the electrons belonging to
the first ensemble, together with the quasi-static tunneling rates. The remaining parameters are the same as in panels (a)--(d) of Fig. 1. The times $t_0$ aregiven in units of the field cycle.}
\end{figure}
 For the
specific parameters of the figure, there exist mainly two sets of electrons
for which the quasi-static rate is large and, therefore, 
whose contributions are relevant: those ejected at $1.5T<t_{0}< 2T$, with
positive momenta, and those released at $2T<t_{0}< 2.5T $, with negative
momenta \cite{footnquasistatic}.

For a large range of absolute phases 
$\phi<\phi_{c}$, 
electron-impact ionization from the latter set of trajectories is classically forbidden. 
 Thus, the distributions concentrate on the first quadrant. 
Around the critical phase, this process is allowed for both sets of electrons
 and the tunneling rates are comparable. Consequently, there are 
relevant contributions to the yield in the first and third quadrants.
 This changes for larger phases, as for instance $\phi=0.9\pi$. In this case, 
the electrons ejected at $2T<t_{0}< 2.5T$ are favored and $p_{j\parallel}$ 
are mainly negative.

\begin{figure}
\begin{center}
\includegraphics[width=4.8cm,angle=270]{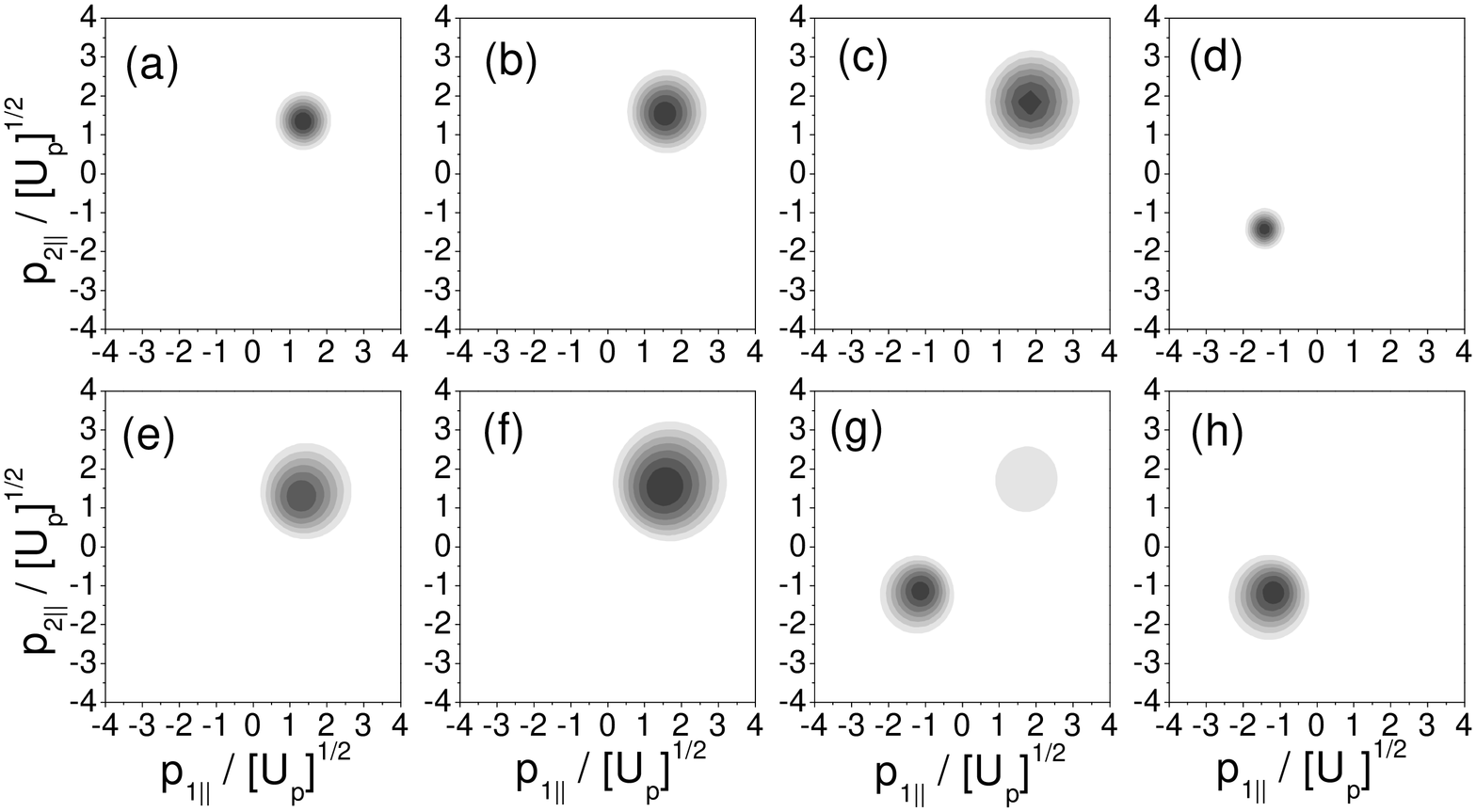}
\end{center}
\caption{Parallel electron momentum distributions for pulses of intensities 
$I=4\times 10^{14}\mathrm{W/cm^2}$ and $I=8\times 10^{14}\mathrm{W/cm^2}$ 
(upper and lower panels, respectively). The electrons were 
ejected with the quasi-static tunneling rate. The remaining parameters are the same as in panels (a)--(d) of Fig. 1.}
\end{figure}
  The role of the phase space is  even more evident as the driving-field
intensity is varied, as shown in Fig. 3.
 By doing so, one is changing the radius of the circle
described by Eq. (\ref{resc2}), and thus the region in the $(p_{1\parallel
},p_{2\parallel })$ plane for which rescattering is
classically allowed.  Therefore, the critical phase may
change.

For a lower intensity than that in Fig. 1, the yield in the
negative momentum region appears for a phase larger than $\phi_c=0.8\pi$ (c.f. panels (c) and (d)). This is due to the fact that the classically allowed
region for $p_{\parallel}<0$ is almost vanishing. Thus, even
if the tunneling rates are comparable,
the first electron, upon return, no longer possesses enough kinetic energy
to release the second electron in a way that both leave with negative parallel momenta. For a higher 
intensity, apart from the fact that both regions are classically allowed, the probability 
that the second electron is 
released with  negative parallel momentum is larger. Therefore, the transition occurs where expected, as displayed in panel (g).  
 This is confirmed by Fig. 4, where, as the intensity decreases,  
the dominant set of ionization times ($2T<t_0<2.5T$), corresponding to $p_{\parallel}<0$, collapses.

\begin{figure}
\begin{center}
\includegraphics[width=4.8cm,angle=270]{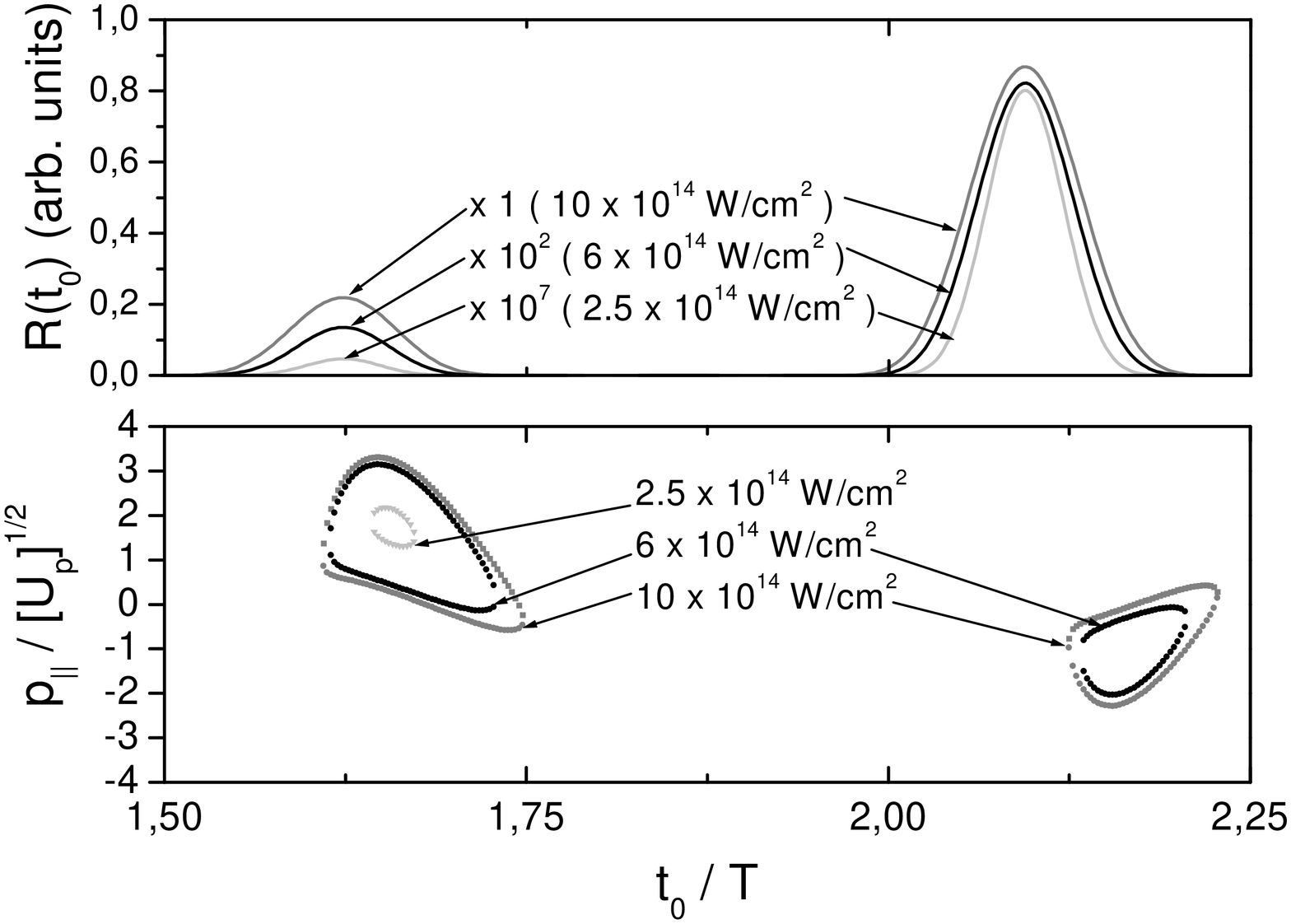}
\end{center}
\caption{Parallel electron momenta $p_{\parallel }$ along $p_{1\parallel
}=p_{2\parallel }$ and for vanishing transverse momenta, as functions of the
start times $t_{0}$, together with
the quasi-static tunneling rates, for 
several driving-field intensities and the critical phase $\phi=0.8\pi$.
The remaining parameters are the same as in Fig. 3.
 For the lowest intensity, 
rescattering caused by electrons ejected at $2T<t_0<2.5T$ is classically forbidden, so that the corresponding curve is not displayed.
}
\end{figure}

In conclusion, we perform a theoretical investigation of
NSDI with few-cycle pulses, using a classical model based
on electron impact ionization.
Both electrons have equal final momentum components 
parallel to the field polarization, which are mainly positive or negative.
 The sign of such momenta and their most probable values 
depend on the absolute phase $\phi$.
In particular, around a critical phase, these
momenta change sign.
Such features are explained 
as the interplay between 
the tunneling rate for the first electron
and the phase space.

 The changes in the NSDI yields upon a critical phase are 
far more 
extreme effects than those observed for
HHG or ATI. In fact, 
 nonsequential double ionization has an advantage over the other two
 phenomena: the phase space region
 contributing to the process is {\it confined}. Thus, for particular 
 ranges of $\phi$, it is easier to make a whole region either
 classically forbidden,
by making the radius in Eq. (6) collapse,  or irrelevant, by reducing the corresponding
ionization rate. 
Therefore, NSDI is, in principle, a tool for absolute phase
diagnosis which is more efficient than the existing schemes.

We thank A. Fring for useful discussions. This work was financed in part by the Deutsche Forschungsgemeinschaft.


\begin{references}
\bibitem{footnati} in ATI, an atom absorbs more photons than it is necessary for it to ionize, releasing high-energy electrons.
\bibitem{footnpond}The ponderomotive energy is the temporal average
$U_p=<A(t)^2>_t/2$, where $A(t)$ is the vector potential. In order to
compare our results more directly with the existing literature, we approximate it
by the expression for the monochromatic case, i.e.,
$U_p=4A_0^2$.
\bibitem{footnint}This only holds if the interaction by which the first
electrons dislodge the second is of contact type,
i.e. $V_{12}=\delta(\mathbf{r}_{1})\delta(\mathbf{r}_2-\mathbf{r}_1)$. This interaction is implicit in Eq. (7). For other types of interactions see, e.g.,
Ref.\cite{preprint0}.
\bibitem{footnspread} Longer excursion times are not relevant due to wave-packet spreading: the overlap between the wave 
packets of the first and of the second electron would be negligible.
In our classical model, this implies leaving the corresponding electrons 
out of the ensemble.
\bibitem{footnquasistatic}
 Other sets of electrons ($t_0\lesssim 1.5T$)
do not contribute to the yield, since the corresponding tunneling rate is very
small. Their influence 
is  
seen if this rate is artificially taken to be constant (c.f. lower panels of Fig.1).
\end{references}
\end{document}